\PassOptionsToPackage{unicode=true}{hyperref} 
\PassOptionsToPackage{hyphens}{url}
\PassOptionsToPackage{dvipsnames,svgnames*,x11names*}{xcolor}
\documentclass[]{article}
\usepackage{lmodern}
\usepackage{amssymb,amsmath}
\usepackage{ifxetex,ifluatex}
\usepackage{fixltx2e} 
\ifnum 0\ifxetex 1\fi\ifluatex 1\fi=0 
  \usepackage[T1]{fontenc}
  \usepackage[utf8]{inputenc}
  \usepackage{textcomp} 
\else 
  \usepackage{unicode-math}
  \defaultfontfeatures{Ligatures=TeX,Scale=MatchLowercase}
\fi
\IfFileExists{upquote.sty}{\usepackage{upquote}}{}
\IfFileExists{microtype.sty}{%
\usepackage[]{microtype}
\UseMicrotypeSet[protrusion]{basicmath} 
}{}
\IfFileExists{parskip.sty}{%
\usepackage{parskip}
}{
\setlength{\parindent}{0pt}
\setlength{\parskip}{6pt plus 2pt minus 1pt}
}
\usepackage{xcolor}
\usepackage{hyperref}
\hypersetup{
            pdftitle={A Simple Algorithm for Estimating Distribution Parameters from n-Dimensional Randomized Binary Responses},
            pdfauthor={Staal A. Vinterbo Department of Information Security and Communication Technology Norwegian University of Science and Technology},
            colorlinks=true,
            linkcolor=Red,
            citecolor=Blue,
            urlcolor=orange,
            breaklinks=true}
\usepackage{color}
\usepackage{fancyvrb}

\DefineVerbatimEnvironment{Highlighting}{Verbatim}{commandchars=\\\{\}}
\newenvironment{Shaded}{}{}

\newcommand{\BuiltInTok}[1]{#1}

\newcommand{\ControlFlowTok}[1]{\textcolor[rgb]{0.00,0.44,0.13}{\textbf{#1}}}

\newcommand{\DecValTok}[1]{\textcolor[rgb]{0.25,0.63,0.44}{#1}}

\newcommand{\ImportTok}[1]{#1}

\newcommand{\KeywordTok}[1]{\textcolor[rgb]{0.00,0.44,0.13}{\textbf{#1}}}
\newcommand{\NormalTok}[1]{#1}
\newcommand{\OperatorTok}[1]{\textcolor[rgb]{0.40,0.40,0.40}{#1}}

\usepackage{graphicx,grffile}
\makeatletter
\def\maxwidth{\ifdim\Gin@nat@width>\linewidth\linewidth\else\Gin@nat@width\fi}
\def\maxheight{\ifdim\Gin@nat@height>\textheight\textheight\else\Gin@nat@height\fi}
\makeatother
\setkeys{Gin}{width=\maxwidth,height=\maxheight,keepaspectratio}
\providecommand{\tightlist}{%
  \setlength{\itemsep}{0pt}\setlength{\parskip}{0pt}}
\ifx\paragraph\undefined\else
\let\oldparagraph\paragraph
\renewcommand{\paragraph}[1]{\oldparagraph{#1}\mbox{}}
\fi
\ifx\subparagraph\undefined\else
\let\oldsubparagraph\subparagraph
\renewcommand{\subparagraph}[1]{\oldsubparagraph{#1}\mbox{}}
\fi

\makeatletter
\def\fps@figure{htbp}
\makeatother

\usepackage[a4paper]{geometry}
\usepackage[small,bf]{caption}
\usepackage{graphicx}
\usepackage{amsmath}
\usepackage{amssymb}
\usepackage{color}
\usepackage{amsthm}

\begingroup
    \makeatletter
    \@for\theoremstyle:=definition,remark,plain\do{%
        \expandafter\g@addto@macro\csname th@\theoremstyle\endcsname{%
            \addtolength\thm@preskip\parskip
            }%
        }
\endgroup

\newtheorem{theorem}{Theorem}[section]
\newtheorem{lemma}[theorem]{Lemma}
\newtheorem{proposition}[theorem]{Proposition}
\newtheorem{corollary}{Corollary}
\theoremstyle{definition}
\newtheorem{definition}{Definition}[section]

\newtheorem{implementation}{Implementation}

\usepackage{cite} 
\let\citep=\cite

\providecommand{\noopsort}[1]{}

\renewcommand\footnotemark{} 

\title{A Simple Algorithm for Estimating Distribution Parameters from
\(n\)-Dimensional Randomized Binary Responses
\thanks{Information Security - 21th International Conference, ISC 2018, preprint.}}
\author{Staal A. Vinterbo \cr\cr Department of Information Security and
Communication Technology \cr Norwegian University of Science and
Technology
\cr\href{mailto://Staal.Vinterbo@ntnu.no}{\footnotesize{Staal.Vinterbo@ntnu.no}}}
\providecommand{\institute}[1]{}
\institute{Department of Information Security and Communication Technology, NTNU}
\date{}

\begin{document}
\maketitle
\begin{abstract}
Randomized response is attractive for privacy preserving data collection
because the provided privacy can be quantified by means such as
differential privacy. However, recovering and analyzing statistics
involving multiple dependent randomized binary attributes can be
difficult, posing a significant barrier to use. In this work, we address
this problem by identifying and analyzing a family of response
randomizers that change each binary attribute independently with the
same probability. Modes of Google's Rappor randomizer as well as
applications of two well-known classical randomized response methods,
Warner's original method and Simmons' unrelated question method, belong
to this family. We show that randomizers in this family transform
multinomial distribution parameters by an iterated Kronecker product of
an invertible and bisymmetric \(2\times 2\) matrix. This allows us to
present a simple and efficient algorithm for obtaining unbiased maximum
likelihood parameter estimates for \(k\)-way marginals from randomized
responses and provide theoretical bounds on the statistical efficiency
achieved. We also describe the efficiency -- differential privacy
tradeoff. Importantly, both randomization of responses and the
estimation algorithm are simple to implement, an aspect critical to
technologies for privacy protection and security.
\end{abstract}

\newcommand{\xqed}[1]{\leavevmode\unskip\penalty9999 \hbox{}\nobreak\hfill\quad\hbox{\ensuremath{#1}}}

\newcommand{\eprop}{}

\newcommand{\eproof}{\leavevmode\unskip\penalty 9999 \hbox{}\nobreak\hfill\quad\hbox{\ensuremath{\square}}}

\newcommand{\mc}[1]{\ensuremath{\mathcal{#1}}}
\newcommand{\mb}[1]{\ensuremath{\mathbf{#1}}}
\newcommand{\mbb}[1]{\ensuremath{\mathbb{#1}}}

\newcommand{\fcomp}{\mathbin{\mathchoice
  {\xcirc\scriptstyle}
  {\xcirc\scriptstyle}
  {\xcirc\scriptscriptstyle}
  {\xcirc\scriptscriptstyle}
}}
\newcommand{\xcirc}[1]{\vcenter{\hbox{$#1\circ$}}}

\newcommand{\cov}{\operatorname{cov}}
\newcommand{\var}{\operatorname{Var}}
\newcommand{\tr}{\operatorname{Tr}}
\newcommand{\E}{\operatorname{E}}

\newcommand{\X}{\ensuremath{\mathcal{X}}}
\newcommand{\Y}{\ensuremath{\mathcal{Y}}}
\newcommand{\strings}{\ensuremath{\mathbb{S}}}
\newcommand{\mult}{\ensuremath{\mathrm{Mult}}}

\newcommand{\bit}{\ensuremath{\mathbb{B}}}
\newcommand{\biti}{\ensuremath{\ensuremath{\mathbb{B}}^\infty}}
\newcommand{\ms}[1]{\ensuremath{\mathit{#1}}}

\newcommand{\had}{\odot}

\newcommand{\loss}{L}

\hypertarget{introduction}{%
\section{Introduction}\label{introduction}}

Randomized response, introduced by Warner in 1965 (Warner 1965), works
by allowing survey respondents to sample their response according to a
particular probability distribution. This provides privacy while still
allowing the surveyor to gain insights about the queried population. Due
to its suitability for large scale privacy preserving data collection,
randomized response has lately enjoyed a resurgence in interest from
Apple and Google, among others.

As an example of randomized response, consider a population of parties
each holding an independent sensitive bit \(b\) with \(P(b = 1) = p\)
where \(p\) is unknown. We want to estimate \(p\) and therefore randomly
select \(m\) parties \(i\) to ask for their bit values \(b_i\) for this
purpose. Tools from information security allow us to collect the bit
values and compute the estimate \(\hat{p} = m^{-1} \sum_i b_i\) of \(p\)
without access to any proper subset sum of bit values. However, if
\(\hat{p} = 1\) we can infer that \(b_i = 1\) for all observed bits.
Even if we are trusted with this knowledge, disseminating \(\hat{p}\)
allows outsiders to infer \(b_i = 1\) for any party \(i\) known to be a
contributor. Knowing this, parties might not be willing to share their
bit-value directly. However, if each contributor is allowed to lie with
a probability \(q < \frac{1}{2}\), then we can argue that
\(\frac{1-q}{q}\) is the upper limit to which any adversary can update
their belief regarding the true value of any contributed bit. If parties
then agree to contribute, we can still estimate \(p\), albeit at a loss
of statistical efficiency.

In 1977, Tore Dalenius defined \emph{disclosure} about an object \(x\)
by a statistic \(v\) with respect to a property \(p\) to have happened
if the value \(p(x)\) can be determined more accurately with knowledge
of \(v\) than without (Dalenius 1977). A goal of information security is
preserving the integrity circles of trust. Mechanisms to achieve this
include access control, communication security, and secure multi-party
computation. These mechanisms have in common that that the protected
information is well circumscribed, and the states of allowed access are
discrete. Disclosure control, on the other hand, provides a tool for
considering questions regarding the \emph{consequence} of access, and
how to deal with information not necessarily well circumscribed. An
emerging standard for defining privacy based on disclosure control is
differential privacy (Dwork et al. 2006). The likelihood ratio
\(\frac{1-q}{q}\) from the example above is an example of a
quantification of disclosure risk, and the log transformation of this
ratio is parameterized in differential privacy. We can also view the
above randomization as enabling continuously graded access to each
contributor's bit, quantifiable by entropy, for example.

In the past, when surveys were conducted manually with responses
recorded on paper, surveys were generally limited to a single randomized
dichotomous question. It was simply too expensive to survey enough
individuals to support efficient recovery of parameter estimates with
multiple randomized questions. With the advent of computerized surveys,
enrollment is much easier, particularly if the data is collected
automatically. With increased enrollment comes the ability to consider
multiple randomized values per respondent and still obtain efficient
estimates of population distribution parameters. On the other hand,
multiple randomized values per response significantly increases the
difficulty of analysis. For example, the first publication regarding
Google's 2014 Rappor technology for automatically collecting end user
data with randomized response (Erlingsson, Pihur, and Korolova 2014)
only considered each bit in an \(n\)-bit response independently as if it
were the only bit randomized. The consideration of several bits jointly,
had to wait for a subsequent publication (Fanti, Pihur, and Erlingsson
2015). Neither of these publications provided theoretical bounds of
efficiency loss due to randomization. The point is that analyzing
multi-question randomized response can be difficult, potentially causing
surveyors to adopt less effective privacy protections.

We address this problem by defining a family of very easily
implementable randomizers of length \(n\)-bit strings or surveys with
\(n\) sensitive dichotomous questions. For the randomizers in this
family, we provide simply computable distribution parameter estimators
as well as statistical efficiency bounds for these. As these randomizers
act on each response bit independently, marginals can be queried and
recovered independently. This is helpful when bit \(k\) to query is
chosen based on the length \(k-1\) based marginal already queried, or
when the bits of responses are distributed among multiple sources.

\hypertarget{contributions-in-detail}{%
\subsection{Contributions in Detail}\label{contributions-in-detail}}

A randomized response method can be seen as a randomized algorithm \(M\)
that takes a response \(x\) as input and produces a randomized response
\(r = M(x)\). Encoding both \(x\) and \(r\) as length \(n\) bit strings,
the algorithm \(M\) can be characterized by a \(2^n \times 2^n\) matrix
\(C\) where the entry indexed by \((r,x)\) contains \(P(r = M(x))\). If
\(M\) is applied independently to each of \(m\) strings sampled
according to a multinomial distribution with parameters
\(m \in \ensuremath{\mathbb{N}}\) and \(\pi \in [0,1]^{2^n}\), the
resulting strings are expected to be multinomially distributed with
parameters \(m\) and \(C\pi\). Consequently, if \(C\) is invertible we
can obtain a maximum likelihood estimate for \(\pi\) from the histogram
\(y\) of observed randomized responses as \(m^{-1}C^{-1}y\). If \(C\) is
not invertible, using the expectation maximization algorithm can be a
suitable, albeit more complicated alternative for obtaining estimates.
In general, the expression of \(C\) can be such that estimators for the
population parameters are not available in closed form (Barabesi,
Franceschi, and Marcheselli 2012).

We first recognize that a randomizer \(M\) that randomizes each bit in a
length \(n\) string \(x\) independently in an identical manner can be
represented by the iterated Kronecker product \(C\) of a bisymmetric
\(2\times 2\) matrix (Theorem \ref{thm:zrm} and Proposition
\ref{prop:zab}). For the family of such randomizers, our contributions
are developments of

\begin{itemize}
\tightlist
\item
  a definition of \(C^{-1}\) in terms of an iterated Kronecker products
  of a bisymmetric \(2\times 2\) matrix,
\item
  closed form formulas for the individual entries of both \(C\) and
  \(C^{-1}\), of which at most \(n+1\) are distinct in each matrix
  (Theorem \ref{thm:cxy} and Corollary \ref{corr:cxy}),
\item
  a closed form formula for the trace of the covariance matrix for the
  unbiased maximum likelihood estimator \(m^{-1} C^{-1} Y\) (Lemma
  \ref{lem:tr}),
\item
  a closed form formula for the effective loss in sample size for
  estimating \(\pi\) due to randomization (Theorem \ref{thm:l2formula})
  together with concentration bounds for uniformly distributed \(\pi\)
  (Proposition \ref{prop:lambda}),
\item
  an analysis of the loss of effective sample size in terms of the
  afforded level of differential privacy (Theorem \ref{thm:dp}).
\end{itemize}

As the Kronecker product can be implemented in linear time in the number
of entries in the result, \(C\) and \(C^{-1}\) for iterated bisymmetric
randomizers can be computed by an algorithm that is linear in the number
of entries of these matrices (Proposition \ref{prop:time}). We show that
this algorithm is simple to state and simple to implement (Section
\ref{sec:alg}), which facilities adoption as well as verification of
implementation correctness. These are both critical aspects of
algorithms applied for privacy protection and security.

Finally, we show that our results apply to modes of Rappor, application
of Warner's original randomizer (Warner 1965) and Simmon's unrelated
question randomizer (Greenberg et al. 1969).

\hypertarget{related-work}{%
\section{Related Work}\label{related-work}}

Randomized response was first introduced primarily as a technique to
reduce bias introduced by absent or untruthful responses to a single
potentially sensitive dichotomous question (Warner 1965). Much research
into randomized response is in the context of an interview tool for
social sciences research. Here, randomization devices generally consist
of a physical source of randomness like a spinner or a coin, together
with a protocol for how the respondent should use it. These devices are
then evaluated in terms of both human factors, e.g., protocol compliance
and response rates, as well as the statistical utility of their
randomized output (Umesh and Peterson 1991; Lensvelt-Mulders et al.
2005). Randomized response surveys carry a double burden of requiring
additional time and effort on behalf of the respondents, as well as
requiring an enrollment that increases rapidly in the number of
questions that require randomization. This might explain why randomized
response designs for single dichotomous sensitive questions (Warner
1965; Greenberg et al. 1969; Folsom et al. 1973; Blair, Imai, and Zhou
2015) are much more common in the literature than for multiple sensitive
questions (Bourke 1982; Barabesi, Franceschi, and Marcheselli 2012) or
polychotomous questions (Abul-Ela, Greenberg, and Horvitz 1967).
Furthermore, multiple authors point out that while there exists a
substantial body of methods research, ``there have been very few
substantive applications {[}of randomized response techniques{]}''
(Blair, Imai, and Zhou 2015; Lensvelt-Mulders et al. 2005; Umesh and
Peterson 1991).

However, as automated data collection on very large populations has
become available, interest in randomized response involving multiple
independent questions has emerged. Two examples are Google's Rappor
(Erlingsson, Pihur, and Korolova 2014) technology for collecting
end-user data, and Apple's technology for collecting analytics data in
MacOS and iOS (Tang et al. 2017; Apple 2017).

The view of randomizers as transformations of multinomial distribution
parameters has been investigated in the context of local differential
privacy (Duchi, Jordan, and Wainwright 2013). Kairouz et al. (Kairouz,
Oh, and Viswanath 2014) analyze what they call staircase mechanisms,
which in the context of this paper can be thought of as family of
randomized response mechanisms where \(C = B D\), where \(B\) is a
matrix that contains at most two values, located on the diagonal and
elsewhere, respectively, and \(D\) is a diagonal matrix. In particular,
they investigate a randomized response mechanism \(k\)-RR where \(D\) is
the identity matrix. For \(k\)-RR they show that this mechanism is
optimal with respect to the tradeoff between differential privacy and
utility defined in terms of KL-divergence. In subsequent work (Kairouz,
Bonawitz, and Ramage 2016) they further analyze Warner's original
proposal, their \(k\)-RR staircase mechanism and Rappor under general
loss functions. They show that for \(n = 1\) Warner's proposal is
optimal for any loss and any differential privacy level. This is the
only case where the \(k\)-RR family of staircase mechanisms intersects
the family of randomizers presented here in this paper. Furthermore,
their analysis is based on entire responses being known up front and it
is not clear how to apply their work in the case where a response is an
interactively queried sequence of \(n > 1\) randomized bits.

\hypertarget{randomizing-mechanisms}{%
\section{Randomizing Mechanisms}\label{randomizing-mechanisms}}

\hypertarget{length-n-bit-strings-and-the-multinomial-distribution}{%
\subsection{\texorpdfstring{Length \(n\) Bit Strings and the Multinomial
Distribution}{Length n Bit Strings and the Multinomial Distribution}}\label{length-n-bit-strings-and-the-multinomial-distribution}}

Let \(\ensuremath{\mathbb{B}}= \{0,1\}\), and let
\(\wedge, \vee, \oplus\) denote logical and, or, and exclusive or. For
\(x,y,u \in \ensuremath{\mathbb{B}}^n\) let
\(x \geq y \iff x \wedge y = y\),
\(x =_u y \iff x \wedge u = y \wedge u\), and
\([x]_u = \{y | x =_u y\}\). Now, let
\(e_i \in \ensuremath{\mathbb{B}}_n\) be the string with a single 1 at
position \(i\), and let for a set of indices \(K\),
\(e_K \geq e_i \iff i \in K\). For
\(x = (x_0, x_1, \ldots, x_{n-1}) \in \ensuremath{\mathbb{B}}^n\),
\(|x| = \sum_{i=0}^{n-1} x_i\). Also let
\(\eta : \ensuremath{\mathbb{B}}^n \to \ensuremath{\mathbb{N}}\) be
defined as
\(\eta(x_0, x_1, \ldots, x_{n-1}) = \sum_{i = 0}^{n-1}x_i2^{i}\), and
let \(\varsigma = n^{-1}\). The function \(\eta\) lets us treat element
\(x \in \ensuremath{\mathbb{B}}^n\) as a 0-based index \(\eta(x)\) which
we will do often. Also, let \(\odot\) denote the coordinate-wise
(Hadamard) product.

Consider an experiment that produces an outcome in
\(\{0, 2, \ldots, k-1\}\) for some positive integer \(k\), where outcome
\(i\) is produced with probability \(p_i\). Let \(m\) indicate a fixed
number of independent experiments and let \(X_i\) denote the number of
times outcome \(i\) is observed among the \(m\) experiments. Note that
\(\sum_{i = 0}^{k-1} X_i = m\). Then
\({X} = (X_0, X_1, \ldots, X_{k-1})\) follows a a multinomial
distribution \(\ensuremath{\mathrm{Mult}}(m,\pi)\) where
\(\pi = (p_0, p_1, \ldots, p_{k-1})\). The variables \({X_i}\) each have
expectation \(mp_i\) and variance \(mp_i(1-p_i)\), and we can think of a
realization of \({X}\) as a histogram over the outcomes of the \(m\)
experiments. Also, when \(m=1\), \({X}\) follows a categorical
distribution, and when \(m=1\) and \(k=2\), \({X}\) follows a Bernoulli
distribution. If experiments produce outcomes in
\(\ensuremath{\mathbb{B}}^n\) for \(n \in \ensuremath{\mathbb{N}}\), we
let \(X_i\) be the number of times
\(\varsigma(i) \in \ensuremath{\mathbb{B}}^n\) is observed among \(m\)
experiments.

The estimator \(\hat{\pi}^*(m) = m^{-1} {X}\) is a maximum likelihood
estimator for \(\pi\) (e.g., (Lehmann and Casella 1998, example 6.11))
and the covariance matrix of \(\hat{\pi}^*(m)\) is \[
\operatorname{cov}(\hat{\pi}^*(m)) = m^{-1} (\text{diag}(\pi \odot(1 - \pi) + \pi \odot\pi) -
\pi \pi^T) = m^{-1} (\text{diag}(\pi) - \pi \pi^T)
\] where \(\text{diag}(v)\) is the square matrix with the elements of
\(v\) along the diagonal.

\hypertarget{randomizers-as-linear-transformations}{%
\subsection{Randomizers as Linear
Transformations}\label{randomizers-as-linear-transformations}}

\label{sec:est}

For a string of bits of length \(n\), we will think of a randomizing
mechanism as a function
\(M : \ensuremath{\mathbb{B}}^n \times \ensuremath{\ensuremath{\mathbb{B}}^\infty}\to \ensuremath{\mathbb{B}}^n \times \ensuremath{\ensuremath{\mathbb{B}}^\infty}\)
that takes a bit string and an infinite sequence of independent
uniformly distributed random bits that serves as the source of
randomness, and returns the randomized string and what remains of the
source of randomness. Since the source of randomness consists of
uniformly and independently distributed bits, we will let the randomness
be implicit and state that randomizer \(M\) is a randomized algorithm
\(M : \ensuremath{\mathbb{B}}^n \to \ensuremath{\mathbb{B}}^n\). We can
define randomizer \(M\) in terms of a \(2^n \times 2^n\) conditional
probability matrix \[
{C_M}_{r,x} = P(r = M(x)).
\] Then, if \(C = C_M\) and \({X}\) is
\(\ensuremath{\mathrm{Mult}}(m, \pi)\), then \({Y} = C {X}\) is
\(\ensuremath{\mathrm{Mult}}(m, C \pi)\). If \(C\) invertible, then we
can let \(\hat{\pi}(m) = m^{-1} C^{-1} {Y}\) be an estimator for \(\pi\)
with \begin{align*}
\operatorname{E}(\hat{\pi}(m)) &= m^{-1} C^{-1} \operatorname{E}({Y}) = m^{-1} C^{-1} m \, C \pi =
\pi,
\end{align*} from which we see that it is unbiased. The invariance
property of maximum likelihood estimators (Casella and Berger 2002,
theorem 7.2.10) states that if \(\hat{\theta}\) is a maximum likelihood
estimator for parameter \(\theta\), then for any function \(g\) we have
that \(g(\hat{\theta})\) is a maximum likelihood estimator for
\(g(\theta)\). Consequently, for invertible matrix \(C\) we have that
\(\hat{\pi}(m) = C^{-1} m^{-1} {{Y}} = m^{-1} C^{-1}{{Y}}\) is an
unbiased maximum likelihood estimator of \(\pi\). The covariance matrix
of \(\hat{\pi}(m)\) is \begin{align*}
\operatorname{cov}(\hat{\pi}(m)) &= \operatorname{cov}(m^{-1} C^{-1} {Y})
= m^{-1} \left(C^{-1} \text{diag}(C\pi) {C^{-1}}^T - \pi \pi^T\right).
\end{align*}

\begin{proposition} \label{prop:l2} If \[
L(m) = \frac{\operatorname{Tr}(\operatorname{cov}(\hat{\pi}(m)))}{\operatorname{Tr}(\operatorname{cov}(\hat{\pi}^*(m)))}
\] where \(\operatorname{Tr}(A)\) denotes the sum of the elements along
the diagonal of square matrix \(A\). Then \(L= L(m) = L(m')\) for any
\(m,m' > 0\), and for \(\alpha \geq 1\), \[
\operatorname{E}\left(\|\hat{\pi}(\alpha Lm) - \pi\|_2^2\right) \leq 
\operatorname{E}\left(\|\hat{\pi}^*(m) - \pi\|_2^2\right).\tag*{}
\]\end{proposition}

By the above proposition, the loss of quality of estimation when using
randomized responses over non-randomized responses can be described by
\(L\).

\hypertarget{bitwise-independent-randomizers}{%
\section{Bitwise Independent
Randomizers}\label{bitwise-independent-randomizers}}

Let \((x:\ensuremath{\mathit{xs}}) \in \ensuremath{\mathbb{B}}^n\) be a
length \(n\) sequence of bits with first bit \(x\) and length \(n-1\)
tail \(\ensuremath{\mathit{xs}}\), and let
\(M = (M' : \ensuremath{\mathit{Ms}})\) be a sequence of bit
randomizers. Now define \begin{align*}
R(\epsilon, x) &= R(M, \epsilon) = \epsilon \\
R(M':\ensuremath{\mathit{Ms}}, x':\ensuremath{\mathit{xs}}) &= M'(x'):R(\ensuremath{\mathit{Ms}}, \ensuremath{\mathit{xs}})
\end{align*} where \(\epsilon\) is the empty sequence. We will think of
\(R\) as a function that maps a length \(n\) sequence of bit randomizers
\(M\) to a randomizer \(R(M)\) of length \(n\) bit strings. We note that
\(R(M)\) randomizes each bit independently, and that any randomizer of
length \(n\) bit strings that randomizes each bit independently can be
written as \(R(M')\) for some sequence of bit randomizers \(M'\). We
will call these bitwise independent randomizers.

We first repeat a result regarding bitwise independent randomizers,
first established by Bourke (Bourke 1982), namely that \(R(M)\) is
defined by the Kronecker product \(\otimes\) of its constituent bit
randomizers.

\begin{theorem}[Bourke 1982] \label{thm:zrm} For a sequence
\(M = (M_0, M_1, \ldots, M_{n-1})\) of independent bit randomizers,
\(C_{R(M)} = C_{M_0} \otimes C_{M_1} \otimes \cdots \otimes C_{M_{n-1}}\).\end{theorem}

We now examine a special case of bitwise independent randomizers in more
detail.

\hypertarget{iterated-bisymmetric-randomizers}{%
\subsection{Iterated Bisymmetric
Randomizers}\label{iterated-bisymmetric-randomizers}}

Let a bisymmetric bit randomizer \(M\) be a bit randomizer that has a
matrix \[
C_M = C_{a,b} = 
\begin{pmatrix}
a & b \\
b & a
\end{pmatrix}
\] that is symmetric about both its main diagonals, i.e., is a
bisymmetric matrix. We now consider bitwise independent randomizers
generated by a sequence of identical bisymmetric bit randomizers.

\begin{definition} \label{def:bbr} The iterated Kronecker product of
bisymmetric \(C_{a,b}\) is \[
C_{a,b}(n) = 
\begin{cases}
C_{a,b} \otimes C_{a,b}(n-1) & \text{if $n > 0$,}\\
(1) & \text{otherwise}. \tag*{}
\end{cases} 
\]\end{definition}

\begin{proposition} \label{prop:zab}Let
\(M = (M_0, M_1, \ldots, M_{n-1})\) be a sequence of identical
bisymmetric bit randomizers with \(C_{M_i} = C_{a,b}\). Then
\(C_{R(M)} = C_{a,b}(n)\). \end{proposition}

The iterated Kronecker product preserves properties of \(C_{a,b}\) in
the sense of the following.

\begin{proposition}

The matrix \(C_{a,b}(n)\)

\begin{enumerate}
\def\labelenumi{\alph{enumi}.}
\tightlist
\item
  is bisymmetric,
\item
  has a constant diagonal, and
\item
  has a constant anti-diagonal, and
\item
  contains at most \(n+1\) distinct entries. 
\end{enumerate}

\label{prop:props}
\end{proposition}

As a consequence we will call \(M\) with \(C_{M} = C_{a,b}(n)\)
\emph{iterated bisymmetric randomizers}. We also note that since a
column in \(C_M\) contains a distribution, the sum of entries must be 1.
Consequently, \(b = 1 - a\), and we can define
\(C_{a}(n) = C_{a, 1 - a}(n)\), and let \(C_M = C_{a}(n)\) for iterated
bisymmetric randomizer \(M\). Our main results regarding iterated
bisymmetric randomizers are the following. First we turn to the results
regarding \(C_M\) and \(C_M^{-1}\).

\begin{theorem} \label{thm:cxy}Let \(M\) be a randomizer with
\(C_M = C_{a}(n)\). Then \[{C_M}_{r,x} = a^{n - d}(1-a)^d\] where
\(d = |r \oplus x|\). If \(a \neq \frac{1}{2}\), then
\[C_M^{-1} = C_{\frac{a}{2a - 1}}(n).\tag*{}\]\end{theorem}

\begin{corollary} \label{corr:cxy} If \(a \neq \frac{1}{2}\), \[
{C_M}^{-1}_{x,r} = \frac{a^{n-d} (a - 1)^d}{(2a - 1)^n} 
\] for \(d = |x \oplus r|\).\end{corollary}

We now apply the above results to determine bounds for the statistical
efficiency of iterated bisymmetric randomizers.

\begin{lemma} \label{lem:tr}Let \(\hat{\pi}\) be the unbiased estimator
defined in Section \ref{sec:est} associated with the randomizer \(M\)
with invertible \(C = C_M = C_{a}(n)\). If \(a \neq \frac 12\) the trace
of the variance-covariance matrix of \(\hat{\pi}\) is given by \[
\operatorname{Tr}(\operatorname{cov}(\hat{\pi})) =  m^{-1}(c - s)
\] where \[
c = \left(\frac{a^2+(1-a)^2}{(2a-1)^2}\right)^n,
\] and \(s = \pi^T\pi = \sum_x p_x^2\). \end{lemma}

\begin{theorem} \label{thm:l2formula}For \(\pi^T\pi < 1\),
\(a \neq 1/2\), and \(\hat{\pi}\) associated with \(M\) as in Lemma
\ref{lem:tr} the loss as defined in Proposition \ref{prop:l2} is given
by \[
L= f_L(s) = \frac{\left(\frac{a^2+(1-a)^2}{(2a-1)^2}\right)^n - s}{1 - s}
\] for \(s = \pi^T\pi = \sum_x p_x^2\). Also,
\(f_L(2^{-n}) \leq L\).\end{theorem}

When \(\pi\) is a random uniformly sampled probability distribution on
\(2^n\) categories\footnote{distributed as the flat Dirichlet
  distribution of order \(2^n\).}, the quantity \(\pi^T\pi\), also known
as the Greenwood statistic, has expected value
\(\operatorname{E}(\pi^T\pi) = \frac{2}{2^n + 1}\) and variance
\(\operatorname{Var}(\pi^T\pi) = \frac{4(2^n-1))}{(2^n+1)^2\;(2^n+2)\;(2^n+3)}\)
(Moran 1947). As \(\pi\) is usually unknown, we can approximate the loss
\(L\) as \(L(n) = f_L(\frac{2}{2^n + 1})\). We state the following about
the quality of this approximation.

\begin{proposition} \label{prop:lambda} For \(n > 2\) and \(\pi\) a
random uniformly sampled probability distribution on \(n\) categories,
\[
P(1 \leq \frac{\operatorname{E}(f_L(\pi^T\pi))}{f_L(\frac{2}{2^n +
1})} \leq 1 + \delta(n)) \geq 0.99.
\] where \(\delta(n) \in O(2^{-3n})\) and \(\delta(3) < 0.2386\).
\end{proposition}

Also, \(\delta(4) < 0.0029\). Finally, we can compare iterated
bisymmetric randomizers in terms of the efficiency of their estimators
\(\hat{\pi}\), which in turn means comparing their associated values for
\(c\) in Lemma \ref{lem:tr}. Here smaller is better.

\hypertarget{a-simple-algorithm}{%
\subsection{A Simple Algorithm}\label{a-simple-algorithm}}

\label{sec:alg}

A randomizer with \(C_M = C_a(n)\) can be implemented as
\(M(x) = x \oplus u\), where \(u\) is a length \(n\) sequence of
independent Bernoulli trials with success probability \(1-a\).

Let \((r_0, r_1, \ldots, r_{m-1})\) be \(m\) \(n\)-bit randomized
responses where each bit has been randomized by an independent
bisymmetric bit randomizer with \(C_M = C_a\), and let \(r_i[K]\) denote
the sub-sequence of \(r_i\) indexed by
\(K \subseteq \{0,1, \ldots, n-1\}\) in order. By Theorem \ref{thm:cxy}
we can then estimate the marginal multinomial distribution parameter
\(\hat{\pi}_K\) for the bits indexed by the \(k\) bits in \(K\) as
follows:

\begin{enumerate}
\def\labelenumi{\arabic{enumi}.}
\tightlist
\item
  let \(y = (y_0, y_1, \ldots, y_{2^k - 1})\) where
  \(y_i = |\{j \mid \eta(r_j[K]) = i \}|\), i.e., \(y\) is the histogram
  over observed sub-sequences, and
\item
  \(\hat{\pi}_K = m^{-1} {C_{\frac{a}{2a - 1}}(k)\; y}.\)
\end{enumerate}

\begin{proposition} \label{prop:time} The simply recursive algorithm
\(C_{a}(n)\) can be implemented to run in time that is linear in the
number of entries of the output matrix.\end{proposition}

A Python code example for implementing the algorithm above is as
follows.

\begin{implementation}

\hspace{0ex}

\footnotesize

\begin{Shaded}
\begin{Highlighting}[]
\ImportTok{from}\NormalTok{ numpy }\ImportTok{import}\NormalTok{ array, kron, log2, bincount }\ImportTok{as}\NormalTok{ bc, arange}

\KeywordTok{def}\NormalTok{ C(n, a):}
\NormalTok{  c, z }\OperatorTok{=}\NormalTok{ array([[a, }\DecValTok{1}\OperatorTok{-}\NormalTok{a], [}\DecValTok{1}\OperatorTok{-}\NormalTok{a, a]]), array([}\DecValTok{1}\NormalTok{])}
  \ControlFlowTok{return}\NormalTok{ z }\ControlFlowTok{if}\NormalTok{ n }\OperatorTok{<} \DecValTok{1} \ControlFlowTok{else}\NormalTok{ kron(c, C(n}\DecValTok{-1}\NormalTok{, a))}

\KeywordTok{def}\NormalTok{ pihat(a, Y):}
\NormalTok{  m, n }\OperatorTok{=} \BuiltInTok{float}\NormalTok{(}\BuiltInTok{sum}\NormalTok{(Y)), }\BuiltInTok{int}\NormalTok{(log2(}\BuiltInTok{len}\NormalTok{(Y)))}
  \ControlFlowTok{return}\NormalTok{ C(n, a}\OperatorTok{/}\NormalTok{(}\DecValTok{2}\OperatorTok{*}\NormalTok{a }\OperatorTok{-} \DecValTok{1}\NormalTok{)).dot(Y)}\OperatorTok{/}\NormalTok{m}

\KeywordTok{def}\NormalTok{ hist(R):}
\NormalTok{  _,n }\OperatorTok{=}\NormalTok{ R.shape}
\NormalTok{  x }\OperatorTok{=} \DecValTok{2}\OperatorTok{**}\NormalTok{(n}\DecValTok{-1}\OperatorTok{-}\NormalTok{arange(n))}
  \ControlFlowTok{return}\NormalTok{ bc(R.dot(x), minlength }\OperatorTok{=} \DecValTok{2}\OperatorTok{**}\NormalTok{n)}
\end{Highlighting}
\end{Shaded}

\normalsize

\label{impl:py}
\end{implementation}

For input \texttt{R} being a \(m \times n\) \texttt{numpy} array of
\(m\) randomized length \(n\) binary responses and \texttt{K} a list of
column indices, the call \texttt{pihat(a,\ hist(R{[}:,K{]}))} computes
the value for \(\hat{\pi}_K\).

\hypertarget{privacy-considerations}{%
\subsection{Privacy Considerations}\label{privacy-considerations}}

\label{sec:privacy}

The results established so far are about efficiency aspects of
estimating multinomial parameters as functions of population and
randomization parameters. We now briefly turn to a measure of privacy
risk for bit-wise independent randomizers.

Now, let \(M\) be a bit randomizer such that entries in \(C_M\) all are
positive, i.e., randomization happens for both possible inputs. We will
in this section only consider such bit randomizers. Let \[
l_M(r) = \frac{\max_x P(r = M(x))}{\min_x P(r = M(x))}.
\] The likelihood ratio \(l_M(r)\) can be thought of representing the
best evidence for preferring one hypothetical input over another when
given the randomized output \(r\). This is reflected in the definition
of Differential Privacy (Dwork et al. 2006), where a randomized
algorithm \(M\) can be considered \(\alpha\)-differentially private if,
for any measurable subset \(S\) of possible outputs, and inputs \(D\)
and \(D'\) obtained from any two sets of individuals that overlap in all
but one individual, we have that \[
\frac{P(M(D) \in S)}{P(M(D') \in S)} \leq \exp(\alpha),
\] and the probabilities are over the randomness available to the
algorithm.

Now let \(l_M = \max_r l_M(r)\), then it follows that \(M\) is a
\(\log(l_M)\)-differentially private algorithm. Now consider \(R(M)\)
for \(n\) bit randomizers \(M = (M_i)_{i = 0}^{n-1}\) such that
\(l_{M_i} \geq l_{M_{i+1}}\), and let \(k \geq |x \oplus x'|\) for any
\(x,x'\) given as input to \(R(M)\). Then
\(l = \prod_{i = 0}^{k-1} l_{M_i}\) can be an upper bound of privacy
loss for any respondent. Specifically, \(R(M)\) is then a
\(\log(l)\)-differentially private algorithm.

Now assume that \(M\) is bisymmetric. Exploiting the structure of \(M\),
we get that \(l_M = l_a = r(a)\) where
\(r(a) = \max(a^{-1}(1-a), a (1 - a)^{-1})\). If \(R(M)\) is an iterated
bisymmetric randomizer, i.e., \(C_{R(M)} = C_{a}(m)\), then
\(l = (l_{a})^k\), and we have that \(R(M)\) is
\(\alpha\)-differentially private for
\(\alpha = \log(l_{a}^k) = k \log(r(a)).\) This is particularly useful
if \(a\) is an invertible function \(a_\phi(\phi)\). Then, we can write
\begin{align*}
\alpha_\phi(\phi) &= k \log(r(a_\phi(\phi))), \text{ and}\\
\phi_\alpha(\alpha) &= a_\phi^{-1}(r^{-1}(\exp(\alpha/k))).
\end{align*} We have from Lemma \ref{lem:tr} that \(c\) is a function
\(c_a(a)\). We can now view \(c\) as a function of \(\phi\) by
\(c_\phi = c_a \mathbin{\mathchoice{\vcenter{\hbox{$\scriptstyle\circ$}}}{\vcenter{\hbox{$\scriptstyle\circ$}}}{\vcenter{\hbox{$\scriptscriptstyle\circ$}}}{\vcenter{\hbox{$\scriptscriptstyle\circ$}}}
}a_\phi\), where
\(\mathbin{\mathchoice{\vcenter{\hbox{$\scriptstyle\circ$}}}{\vcenter{\hbox{$\scriptstyle\circ$}}}{\vcenter{\hbox{$\scriptscriptstyle\circ$}}}{\vcenter{\hbox{$\scriptscriptstyle\circ$}}}
}\) denotes function composition. Expanding further, we can let \(c\) be
a function of \(\alpha\) as
\(c_\alpha = c_\phi \mathbin{\mathchoice{\vcenter{\hbox{$\scriptstyle\circ$}}}{\vcenter{\hbox{$\scriptstyle\circ$}}}{\vcenter{\hbox{$\scriptscriptstyle\circ$}}}{\vcenter{\hbox{$\scriptscriptstyle\circ$}}}
}\phi_\alpha\). This means that \begin{align*}
c_\alpha &= c_a \mathbin{\mathchoice{\vcenter{\hbox{$\scriptstyle\circ$}}}{\vcenter{\hbox{$\scriptstyle\circ$}}}{\vcenter{\hbox{$\scriptscriptstyle\circ$}}}{\vcenter{\hbox{$\scriptscriptstyle\circ$}}}
}a_\phi \mathbin{\mathchoice{\vcenter{\hbox{$\scriptstyle\circ$}}}{\vcenter{\hbox{$\scriptstyle\circ$}}}{\vcenter{\hbox{$\scriptscriptstyle\circ$}}}{\vcenter{\hbox{$\scriptscriptstyle\circ$}}}
}\phi_\alpha \\
&= c_a \mathbin{\mathchoice{\vcenter{\hbox{$\scriptstyle\circ$}}}{\vcenter{\hbox{$\scriptstyle\circ$}}}{\vcenter{\hbox{$\scriptscriptstyle\circ$}}}{\vcenter{\hbox{$\scriptscriptstyle\circ$}}}
}a_\phi \mathbin{\mathchoice{\vcenter{\hbox{$\scriptstyle\circ$}}}{\vcenter{\hbox{$\scriptstyle\circ$}}}{\vcenter{\hbox{$\scriptscriptstyle\circ$}}}{\vcenter{\hbox{$\scriptscriptstyle\circ$}}}
}a_\phi^{-1} \mathbin{\mathchoice{\vcenter{\hbox{$\scriptstyle\circ$}}}{\vcenter{\hbox{$\scriptstyle\circ$}}}{\vcenter{\hbox{$\scriptscriptstyle\circ$}}}{\vcenter{\hbox{$\scriptscriptstyle\circ$}}}
}r^{-1} \mathbin{\mathchoice{\vcenter{\hbox{$\scriptstyle\circ$}}}{\vcenter{\hbox{$\scriptstyle\circ$}}}{\vcenter{\hbox{$\scriptscriptstyle\circ$}}}{\vcenter{\hbox{$\scriptscriptstyle\circ$}}}
}\exp{} \mathbin{\mathchoice{\vcenter{\hbox{$\scriptstyle\circ$}}}{\vcenter{\hbox{$\scriptstyle\circ$}}}{\vcenter{\hbox{$\scriptscriptstyle\circ$}}}{\vcenter{\hbox{$\scriptscriptstyle\circ$}}}
}(x
\mapsto x/k)\\
&= c_a \mathbin{\mathchoice{\vcenter{\hbox{$\scriptstyle\circ$}}}{\vcenter{\hbox{$\scriptstyle\circ$}}}{\vcenter{\hbox{$\scriptscriptstyle\circ$}}}{\vcenter{\hbox{$\scriptscriptstyle\circ$}}}
}r^{-1} \mathbin{\mathchoice{\vcenter{\hbox{$\scriptstyle\circ$}}}{\vcenter{\hbox{$\scriptstyle\circ$}}}{\vcenter{\hbox{$\scriptscriptstyle\circ$}}}{\vcenter{\hbox{$\scriptscriptstyle\circ$}}}
}\exp{} \mathbin{\mathchoice{\vcenter{\hbox{$\scriptstyle\circ$}}}{\vcenter{\hbox{$\scriptstyle\circ$}}}{\vcenter{\hbox{$\scriptscriptstyle\circ$}}}{\vcenter{\hbox{$\scriptscriptstyle\circ$}}}
}(x \mapsto x/k).
\end{align*} The last equation shows that \(c_\alpha\) is independent of
the functional shape of \(a(\phi)\), and therefore this holds for \(L\)
as well. In other words, from a perspective of differential privacy, the
performance of iterated bisymmetric randomizers in terms of \(L\) is
independent of the functional shape of invertible \(a(\phi)\). The above
reasoning proves the following Theorem.

\begin{theorem} \label{thm:dp} For any \(\alpha\)-differentially private
iterated bisymmetric randomizer for inputs not differing in more than
\(k\) bits, for \(c\) from Lemma \ref{lem:tr}, \[
c \geq c_\alpha(\alpha) =
\left(\frac{\exp(2\alpha/k)+1)}{(\exp(\alpha/k)-1)^2}\right)^n,
\] and for \(L\) from Theorem \ref{thm:l2formula}, \[
L\geq L(\alpha) = \frac{c_\alpha(\alpha) - s}{1-s}
\] where \(s = \pi^T\pi\). \end{theorem}

Combining the above with Theorem 2 in (Kairouz, Bonawitz, and Ramage
2016), stating the optimality of Warner's randomizer with regards to the
privacy-utility tradeoff for any loss function and privacy level, we
conclude the following.

\begin{corollary} \label{corr:dp}For \(n=1\), bisymmetric iterated
randomizers are optimal with respect to loss \(L\) for any privacy level
\(\alpha\). \end{corollary}

\hypertarget{case-studies}{%
\section{Case Studies}\label{case-studies}}

\label{sec:study}

\hypertarget{the-unrelated-uniform-question-device}{%
\subsection{The Unrelated Uniform Question
Device}\label{the-unrelated-uniform-question-device}}

In Simmons' unrelated question method, the interviewer asks the
respondent to answer a question randomly selected between the sensitive
question of interest and an unrelated question that the respondent
presumably has no problem answering truthfully. The unrelated question
is chosen with probability \(p\). Here we analyze the variant of this
method where the unrelated question is ``Flip a coin. Is it heads?''. In
other words, the case where interviewer knows that the answer to the
unrelated question is uniformly distributed in the study sample.

Let \(A : \{0,1\}^n \times \{0,1\}^n \times\{0,1\}^n \to \{0,1\}^n\) be
given by \[
A(x, u, z) = x \odot(\ensuremath{\mathbf{1}}-u) + z \odot u
\] where \(\ensuremath{\mathbf{1}}\) is the string with all elements 1.
Letting \(B^n_p\) denote a sequence of \(n\) independent Bernoulli
variables each taking value 1 with probability \(p\), the randomizer
\(M(x)\) can then be defined as \(M(x) = A(x, u, z)\) where \(u\) and
\(z\) are realizations of \(B^n_p\) and \(B^n_{0.5}\) variables,
respectively.

\begin{figure}
\centering
\includegraphics[width=0.99\textwidth,height=\textheight]{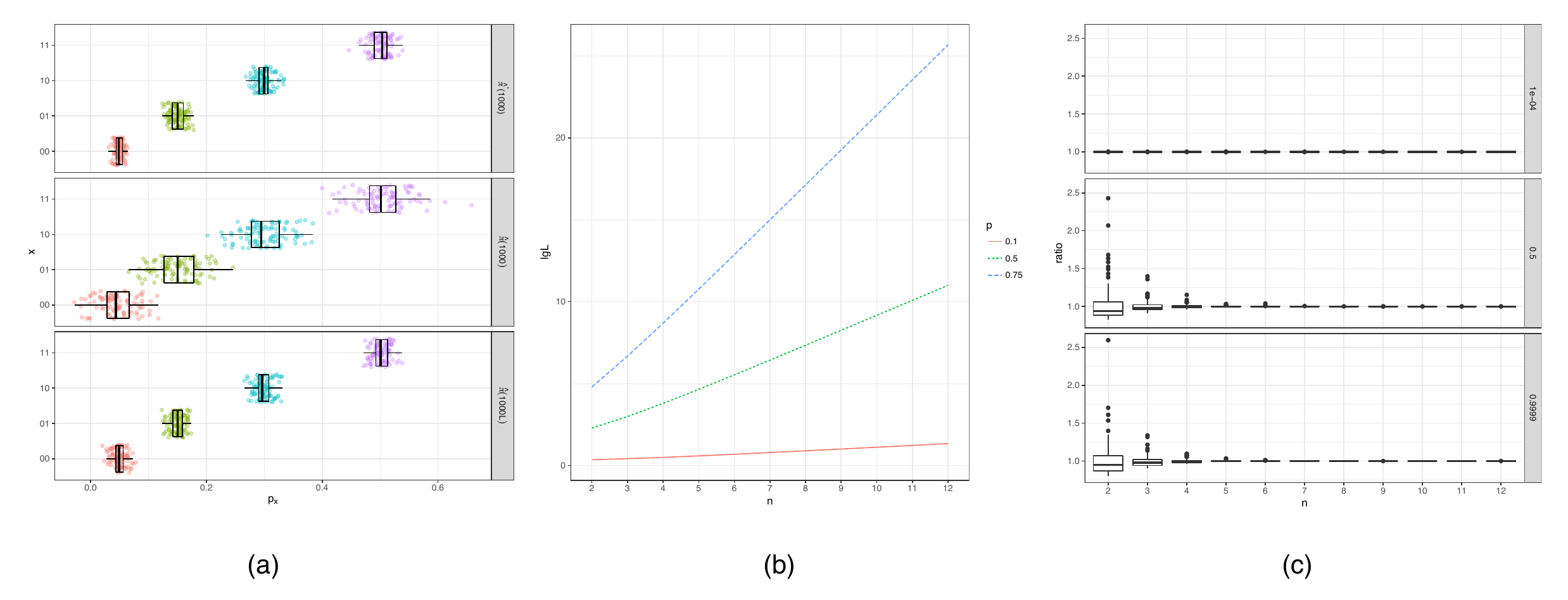}
\caption{(a): Plot of \(\hat{\pi}^*(1000)\), \(\hat{\pi}(1000)\), and
\(\hat{\pi}(L1000)\) for 100 randomly generated datasets with the same
fixed \(\pi = (0.05, 0.15, 0.3, 0.5)^T\), \(p=0.5\), and \(L= 9.75\).
(b): \(\log(L) = \log(f_L(2(2^n + 1)^{-1}))\) for
\(n = 1, 2, \ldots, 12\) and three values of \(p\). (c): The ratio
\(\frac{f_L(\pi^T\pi)}{f_L(2(2^n + 1)^{-1})}\) for 100 random uniformly
sampled \(\pi\) for each \(2 \leq n \leq 12\) and three values of \(p\)
\label{fig:3fig}}
\end{figure}

We start by noting that if \(A(x, u, z) = r\), then
\(u \geq d = x \oplus r\). Now, let \(U\), and \(Z\) be independent
\(B^n_p\) and \(B^n_{0.5}\) variables, respectively. Then \[
P(r = A(x, U, Z)) = \sum_{u \geq d} P(U = u)P(Z =_u r),
\] were \(d = r \oplus x\). Now, \(P(U = u) = p^{|u|}(1 - p)^{n - |u|}\)
and \(P(Z =_u r) = (1/2)^{|u|}\). Furthermore, there are \(2^{n - |d|}\)
strings \(u\) such that \(u \geq d\), and of those
\(\binom{n - |d|}{i}\) have \(|u| = i + |d|\). Consequently,
\begin{align*}
c_{r,x} &= P(r = A(x, U, Z)) 
= \sum_{i = 0}^{n - |d|}
\binom{n - |d|}{i} \left(\frac{p}{2}\right)^{i + |d|} (1 - p)^{n - (i + |d|)}.
\end{align*} If we instead note that each bit is randomized
independently by a bit randomizer \(M_S\) with matrix
\(C_{\frac{2 - p}{2}}\), then by applying Theorem \ref{thm:cxy} we get
that for \(n \geq 1\) \[
{C_M}_{r,x} = \frac{{{p}^{|x \oplus r|}}{{\left( 2-p\right) }^{n-|x \oplus
r|}} }{{{2}^{n}}}.
\] Algebraic manipulations yield that \(c_{r,x} = {C_M}_{r,x}\), and the
value for \(c\) in Lemma \ref{lem:tr} is \[
c_M = \left(\frac{p^2-2p+2}{2(p-1)^2}\right)^n.
\] Figure \ref{fig:3fig}(a) shows the effect of increasing the number of
randomized response data points by a factor \(L\) for computing
\(\hat{\pi}\). As expected, the plot for \(\hat{\pi}(L1000)\) is close
to the target \(\hat{\pi}^*(1000)\). Figure \ref{fig:3fig}(b) shows the
growth of \(\log(f_L(2 (2^n + 1)^{-1}))\) in \(n\) for three values of
\(p\). Figure \ref{fig:3fig}(c) plots the ratio of \(L\) and the
approximated loss \(f_L(2 (2^n + 1)^{-1})\) for 100 uniformly random
distributions \(\pi\) and three probabilities \(p\) of 0.0001, 0.5, and
0.9999.

\hypertarget{warners-original-device-a-randomizer-comparison}{%
\subsection{Warner's Original Device: a Randomizer
Comparison}\label{warners-original-device-a-randomizer-comparison}}

\begin{figure}
\centering
\includegraphics[width=0.8\textwidth,height=\textheight]{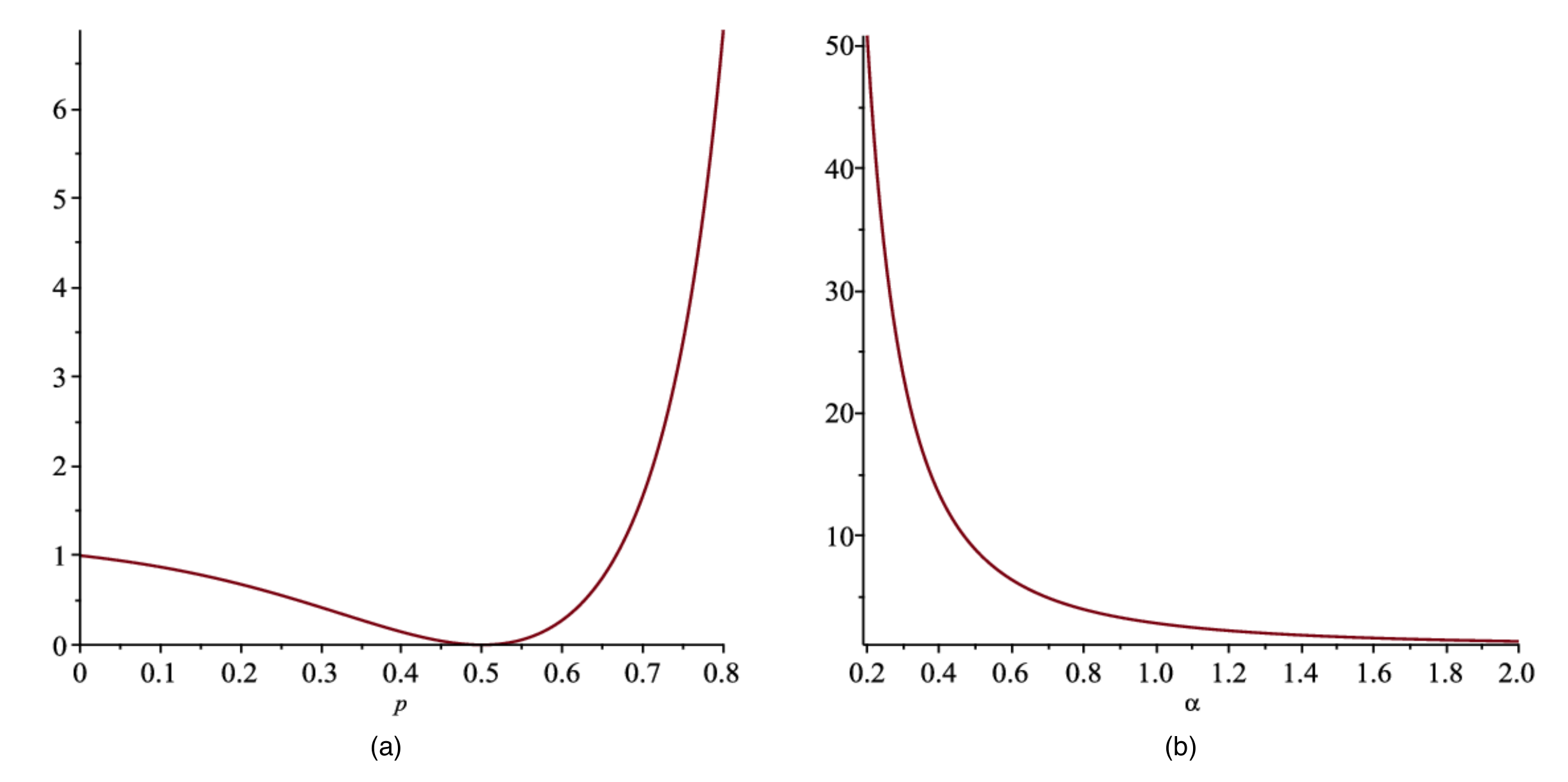}
\caption{(a) The ratio \({c_M}/{c_W}\) for \(p \in (0,0.8)\) and
\(n=1\). (b) \(c_M(\alpha) = c_W(\alpha)\) for \(n=1\) and
\(\alpha \in [0.2,2]\)\label{fig:c1c2}}
\end{figure}

Warner's original randomizer involved a spinner with two areas ``Yes''
and ``No'', with a probability \(p\) for indicating ``Yes''. The
respondent was then asked to spin the spinner unseen by the interviewer
and respond with ``yes'' if the spinner indicated the respondent's true
answer to the sensitive question and ``no'' otherwise. The corresponding
bit randomizer \(M_W\) has matrix \(C_p\), which is invertible if
\(p \neq 0.5\). Furthermore, we have that the value for \(c\) as defined
in Lemma \ref{lem:tr} is \[
c_{W} = \left(\frac{2p^2-2p+1}{(2p-1)^2}\right)^n.
\] We can use the ratio \(c_M/c_{W}\) to compare the estimators for
\(\hat{\pi}\) corresponding to the independent question and Warner's
original method, respectively. We have that \({c_M}/{c_{W}} < 1\) for
\(p \in (0,\frac{2}{3})\). Figure \ref{fig:c1c2} (a) shows a plot of
this ratio for \(n=1\). We see that when \(p < \frac{2}{3}\), the
unrelated question randomizer is preferable with respect to estimating
\(\pi\), and particularly so around \(p = 0.5\) (\(c_{{W}}\) is
undefined at \(p=0.5\)). The preference regions are emphasized as \(n\)
increases. However, if we express \(c_M\) and \(c_W\) as functions of
privacy level \(\alpha\) using the results from Section
\ref{sec:privacy}, we get that the two randomizers perform identically
as \[
c_M(\alpha) = c_W(\alpha) = 
\left(\frac{\exp(2\alpha)+1}{(\exp(\alpha)-1)^2}\right)^n.
\] Figure \ref{fig:c1c2} (b) shows \(c_M(\alpha) = c_W(\alpha)\) for
\(\alpha \in [0.2, 2]\) and \(n=1\).

\hypertarget{the-rappor-randomizer}{%
\subsection{The Rappor Randomizer}\label{the-rappor-randomizer}}

In Rappor, randomization is applied after an hashing of ordinal values
onto a bit string. Here we only examine the randomizer.

The Rappor randomizer is a bit-wise independent randomizer
\(R(M_1, \ldots, M_n)\) where \(M_i = M\) for all \(i\). The bit
randomizer \(M\) is a combination of two bit randomizers, ``Permanent
Randomized Response'' (PR) and ``Instantaneous Randomized Response''
(IR), respectively. The PR randomizer is the above \(M_S\) randomizer
with \(p=f\), while \[
C_{M_{\mathrm{IR}}} = \begin{pmatrix}
1-p & 1-q \\
p & q
\end{pmatrix}.
\] A bit \(b\) decides the combination, where \(b=1\) is called
``one-time'' mode, and
\(M(b) = (1-b)(M_{\mathrm{IR}} \mathbin{\mathchoice{\vcenter{\hbox{$\scriptstyle\circ$}}}{\vcenter{\hbox{$\scriptstyle\circ$}}}{\vcenter{\hbox{$\scriptscriptstyle\circ$}}}{\vcenter{\hbox{$\scriptscriptstyle\circ$}}}
}M_S) + b\,M_{S}\). Consequently, \(C_{M(1)} = C_{\frac{2 - f}{2}}\).
When \(p = 1 - q\), we recognize \(M_{\mathrm{IR}}\) as Warner's \(M_W\)
with parameter \(q\), and
\(C_{M(0)} = C_{q} \, C_{\frac{2 - f}{2}} = C_{q - (q-\frac{1}{2})f}\).
This means that the Rappor randomizer is an iterated bisymmetric
randomizer when \(p = 1 - q\) or \(b=1\).

\hypertarget{conclusion}{%
\section{Conclusion}\label{conclusion}}

A family of randomized response methods is described and analyzed.
Instances of both well known classical and recently developed methods
belong to this family. The analysis resulted in an efficient algorithm
for estimating multinomial population parameters from randomized
responses, and the statistical efficiency of the produced estimates was
described.

The investigated statistical loss grows exponentially in the number of
dimensions \(n\), as does the size of the matrix \(C\) that describes
the effect of randomization on the multinomial parameters. Consequently,
the estimation of these multinomial parameters is only practical for
small \(n\), even with a large number of observations. However, the
knowledge of how statistical loss grows with dimensionality allows the
determination of a value \(k\) for which it is feasible to estimate
parameters for size \(k\) marginals. Since individual variables are
randomized independently, only the variables in the relevant feasible
marginals need to be obtained. Furthermore, due to the independent
randomization of variables these can be queried interactively across
distributed data sources.

Brevity is said to be a hallmark of simplicity (SOSA 2018). Simple
algorithms are more likely to be implemented and trusted by
practitioners, their implementations are easier to maintain and adapt to
changing contexts, and they are easier to implement in constrained
environments such as in hardware (Muller-Hannemann and Schirra 2010).
Simple algorithms are also easier to debug and implement correctly,
which is critical in systems that need to implement privacy and security
requirements. The algorithm presented here is simple. It centers on a
short recursive definition of the matrix \(C^{-1}\), which is shown
implemented in four lines of Python, a multi-purpose programming
language with a significant current market-share. Furthermore, an
implementation of the full process of computing parameter estimates from
binary randomized input was implemented in an additional nine lines of
Python code, making iterated bisymmetric randomizers a potentially
attractive alternative for randomized response applications.

\subsubsection*{Acknowledgments}

Thanks go to the anonymous reviewers for their comments. This work was
in part funded by Oppland fylkeskommune.

\hypertarget{references}{%
\section{References}\label{references}}

\hypertarget{refs}{}
\leavevmode\hypertarget{ref-abul-ela_multi-proportions_1967}{}%
Abul-Ela, Abdel-Latif A., Bernard G. Greenberg, and Daniel G. Horvitz.
1967. ``A Multi-Proportions Randomized Response Model.'' \emph{Journal
of the American Statistical Association} 62 (319): 990--1008.
\url{https://doi.org/10.2307/2283687}.

\leavevmode\hypertarget{ref-apple_learning_2017}{}%
Apple. 2017. ``Learning with Privacy at Scale - Apple.'' \emph{Apple
Machine Learning Journal}.
\url{https://machinelearning.apple.com/2017/12/06/learning-with-privacy-at-scale.html}.

\leavevmode\hypertarget{ref-barabesi_randomized_2012}{}%
Barabesi, Lucio, Sara Franceschi, and Marzia Marcheselli. 2012. ``A
Randomized Response Procedure for Multiple-Sensitive Questions.''
\emph{Stat Papers} 53 (3): 703--18.
\url{https://doi.org/10.1007/s00362-011-0374-5}.

\leavevmode\hypertarget{ref-blair_design_2015}{}%
Blair, Graeme, Kosuke Imai, and Yang-Yang Zhou. 2015. ``Design and
Analysis of the Randomized Response Technique.'' \emph{Journal of the
American Statistical Association} 110 (511): 1304--19.
\url{https://doi.org/10.1080/01621459.2015.1050028}.

\leavevmode\hypertarget{ref-bourke_randomized_1982}{}%
Bourke, Patrick D. 1982. ``Randomized Response Multivariate Designs for
Categorical Data.'' \emph{Communications in Statistics - Theory and
Methods} 11 (25): 2889--2901.
\url{https://doi.org/10.1080/03610928208828430}.

\leavevmode\hypertarget{ref-casella_statistical_2002}{}%
Casella, George., and Roger L. Berger. 2002. \emph{Statistical
Inference}. Australia; Pacific Grove, CA: Duxbury/Thomson Learning.

\leavevmode\hypertarget{ref-dalenius_towards_1977}{}%
Dalenius, Tore. 1977. ``Towards a Methodology for Statistical Disclosure
Control.'' \emph{Statistisk Tidskrift} 15 (429-444): 2--1.

\leavevmode\hypertarget{ref-duchi_local_2013}{}%
Duchi, J. C., M. I. Jordan, and M. J. Wainwright. 2013. ``Local Privacy
and Statistical Minimax Rates.'' In \emph{2013 51st Annual Allerton
Conference on Communication, Control, and Computing (Allerton)},
1592--2. \url{https://doi.org/10.1109/Allerton.2013.6736718}.

\leavevmode\hypertarget{ref-Dwork2006a}{}%
Dwork, Cynthia, Frank McSherry, Kobbi Nissim, and Adam Smith. 2006.
``Calibrating Noise to Sensitivity in Private Data Analysis.'' In
\emph{Proceedings of the Conference on Theory of Cryptography}.
\url{https://doi.org/10.1007/11681878/\%5F14}.

\leavevmode\hypertarget{ref-erlingsson_rappor_2014}{}%
Erlingsson, Úlfar, Vasyl Pihur, and Aleksandra Korolova. 2014. ``RAPPOR:
Randomized Aggregatable Privacy-Preserving Ordinal Response.'' In
\emph{Proceedings of the 2014 ACM SIGSAC Conference on Computer and
Communications Security}, 1054--67. CCS '14. New York, NY, USA: ACM.
\url{https://doi.org/10.1145/2660267.2660348}.

\leavevmode\hypertarget{ref-fanti_building_2015}{}%
Fanti, Giulia, Vasyl Pihur, and Úlfar Erlingsson. 2015. ``Building a
RAPPOR with the Unknown: Privacy-Preserving Learning of Associations and
Data Dictionaries.'' \emph{arXiv:1503.01214 {[}Cs{]}}, March.
\url{http://arxiv.org/abs/1503.01214}.

\leavevmode\hypertarget{ref-folsom_two_1973}{}%
Folsom, Ralph E., Bernard G. Greenberg, Daniel G. Horvitz, and James R.
Abernathy. 1973. ``The Two Alternate Questions Randomized Response Model
for Human Surveys.'' \emph{Journal of the American Statistical
Association} 68 (343): 525--30. \url{https://doi.org/10.2307/2284771}.

\leavevmode\hypertarget{ref-greenberg_unrelated_1969}{}%
Greenberg, Bernard G., Abdel-Latif A. Abul-Ela, Walt R. Simmons, and
Daniel G. Horvitz. 1969. ``The Unrelated Question Randomized Response
Model: Theoretical Framework.'' \emph{Journal of the American
Statistical Association} 64 (326): 520--39.
\url{https://doi.org/10.2307/2283636}.

\leavevmode\hypertarget{ref-kairouz_discrete_2016}{}%
Kairouz, Peter, Keith Bonawitz, and Daniel Ramage. 2016. ``Discrete
Distribution Estimation Under Local Privacy.'' In \emph{Proceedings of
the 33rd International Conference on International Conference on Machine
Learning - Volume 48}, 2436--44. ICML'16. New York, NY, USA: JMLR.org.
\url{http://dl.acm.org/citation.cfm?id=3045390.3045647}.

\leavevmode\hypertarget{ref-kairouz_extremal_2014-1}{}%
Kairouz, Peter, Sewoong Oh, and Pramod Viswanath. 2014. ``Extremal
Mechanisms for Local Differential Privacy.'' In \emph{Advances in Neural
Information Processing Systems 27}, edited by Z. Ghahramani, M. Welling,
C. Cortes, N. D. Lawrence, and K. Q. Weinberger, 2879--87. Curran
Associates, Inc.
\url{http://papers.nips.cc/paper/5392-extremal-mechanisms-for-local-differential-privacy.pdf}.

\leavevmode\hypertarget{ref-lehmann_theory_1998}{}%
Lehmann, E. L., and G. Casella. 1998. \emph{Theory of Point Estimation.}
Springer.

\leavevmode\hypertarget{ref-lensvelt-mulders_meta-analysis_2005}{}%
Lensvelt-Mulders, Gerty J. L. M., Joop J. Hox, Peter G. M.
\noopsort{heijden}van der Heijden, and Cora J. M. Maas. 2005.
``Meta-Analysis of Randomized Response Research: Thirty-Five Years of
Validation.'' \emph{Sociological Methods \& Research} 33 (3): 319--48.
\url{https://doi.org/10.1177/0049124104268664}.

\leavevmode\hypertarget{ref-moran_random_1947-1}{}%
Moran, P. A. P. 1947. ``The Random Division of an Interval.''
\emph{Supplement to the Journal of the Royal Statistical Society} 9 (1):
92--98. \url{https://doi.org/10.2307/2983572}.

\leavevmode\hypertarget{ref-Muller-HannemannAlgorithmEngineeringBridging2010}{}%
Muller-Hannemann, Matthias, and Stefan Schirra, eds. 2010.
\emph{Algorithm Engineering: Bridging the Gap Between Algorithm Theory
and Practice}. Berlin, Heidelberg: Springer-Verlag.

\leavevmode\hypertarget{ref-sosa_symposium_2018}{}%
SOSA. 2018. ``Symposium on Simplicity in Algorithms.'' \emph{SOSA}.
\url{https://simplicityalgorithms.wixsite.com/sosa/cfp}.

\leavevmode\hypertarget{ref-tang_privacy_2017}{}%
Tang, Jun, Aleksandra Korolova, Xiaolong Bai, Xueqiang Wang, and
Xiaofeng Wang. 2017. ``Privacy Loss in Apple's Implementation of
Differential Privacy on MacOS 10.12.'' \emph{arXiv:1709.02753 {[}Cs{]}},
September. \url{http://arxiv.org/abs/1709.02753}.

\leavevmode\hypertarget{ref-umesh_critical_1991}{}%
Umesh, U. N., and Robert A. Peterson. 1991. ``A Critical Evaluation of
the Randomized Response Method: Applications, Validation, and Research
Agenda.'' \emph{Sociological Methods \& Research} 20 (1): 104--38.
\url{https://doi.org/10.1177/0049124191020001004}.

\leavevmode\hypertarget{ref-warner_randomized_1965}{}%
Warner, Stanley L. 1965. ``Randomized Response: A Survey Technique for
Eliminating Evasive Answer Bias.'' \emph{Journal of the American
Statistical Association} 60 (309): 63--69.
\url{https://doi.org/10.1080/01621459.1965.10480775}.

\bibliography{bibliography}

\appendix

\hypertarget{proofs}{%
\section{Proofs}\label{proofs}}

\label{sec:proofs}

We start by making a key observation.

\subsubsection*{Observation 1:}

Consider the \(2^n \times 2^n\) matrix \(C\). If we let entry
\(C_{ix', jy'} = \eta((ix') \oplus (jy')) = 2^{\eta(i \oplus j)} \;\eta(x' \oplus y'),\)
we get that \(C_{x,y} = 1^{n - |x \oplus y|} \; 2^{|x \oplus y|}\).
Since we can write \(C = J_1 \otimes J_2 \otimes \cdots \otimes J_n\)
where \(J_k\) is the \(2 \times 2\) matrix \(J\) such that
\(J_{i,j} = 2^{i \oplus j}\), i.e.,
\(J = \begin{pmatrix} 1 & 2 \\ 2 & 1 \end{pmatrix},\) we get that
\(D_{x,y} = a^{n - |x \oplus y|} \; b^{|x \oplus y|}\) for
\(D = C_{a,b}(n) = B_1 \otimes B_2 \otimes \cdots \otimes B_n\) where
\(B_k = \begin{pmatrix} a & b \\ b & a \end{pmatrix}.\)

\subsubsection*{Proof of Proposition \ref{prop:l2}:}

Note that we can write \begin{align*}
&\operatorname{Tr}(\operatorname{cov}(\hat{\pi}(m))) = \sum_x \operatorname{Var}(\hat{\pi}_x(m)), \text{ }
\operatorname{Tr}(\operatorname{cov}(\hat{\pi}^*(m))) = \sum_x \operatorname{Var}(\hat{\pi}_x^*(m))\\
&\operatorname{Var}(\hat{\pi}_x(m)) = m^{-1} F(x,\pi), \text{ }
\operatorname{Var}(\hat{\pi}^*_x(m)) = m^{-1} G(x,\pi)
\end{align*} where \(F\) and \(G\) are functions independent of \(m\).
Then for any positive integer \(m\), \[
L(m) = \frac{m^{-1} \sum_x F(x,\pi)}{m^{-1} \sum_x G(x,\pi)}
    = \frac{\sum_x F(x,\pi)}{\sum_x G(x,\pi)} = L.
\]\\
and \begin{align*}
\sum_x \operatorname{Var}(\hat{\pi}_x(\alpha Lm)) 
&= \sum_x \alpha^{-1}  m^{-1} L^{-1} F(x,\pi)  
\leq m^{-1} L^{-1} \sum_x F(x, \pi)\\
&= \sum_x m^{-1} G(x, \pi) 
= \sum_x \operatorname{Var}(\hat{\pi}^*_x(m)).
\end{align*} Furthermore, \begin{align*}
\operatorname{E}\left(\|\hat{\pi}(m) - \pi\|_2^2\right) &= 
\operatorname{E}\left(\sum_x (\hat{\pi}_x(m) - p_x)^2\right)  
= \sum_x \operatorname{E}\left((\hat{\pi}_x(m) - p_x)^2\right) \\
&= \sum_x \operatorname{Var}(\hat{\pi}_x(m)),
\end{align*} and similarly
\(\operatorname{E}\left(\|\hat{\pi}^*(m) - \pi\|_2^2\right) = \sum_x\operatorname{Var}(\hat{\pi}^*_x(m))\).\leavevmode\unskip\penalty 9999 \hbox{}\nobreak\hfill\quad\hbox{\ensuremath{\square}}

\subsubsection*{Proof of Proposition \ref{prop:zab}:}

The proposition follows directly from Theorem \ref{thm:zrm}.
\leavevmode\unskip\penalty 9999 \hbox{}\nobreak\hfill\quad\hbox{\ensuremath{\square}}

\subsubsection*{Proof of Proposition \ref{prop:props}:}

We first note that for \(i \in \{0, 1, \ldots, 2^n - 1\}\) we have that
\(\varsigma((2^n - 1) - i) = \ensuremath{\mathbf{1}} \oplus \varsigma(i)\).
From this and that \(\oplus\) commutes, we get

\begin{enumerate}
\def\labelenumi{\arabic{enumi}.}
\tightlist
\item
  \(|\varsigma(i) \oplus \varsigma(n - i)| = n\), and
\item
  \(|\varsigma(i) \oplus \varsigma(j)| = |\varsigma(2^n - 1 - j) \oplus \varsigma(2^n - 1 - i)|\).
\end{enumerate}

The above and that the entry
\(C_{a,b}(n)_{i,j} = g(|\varsigma(i) \oplus \varsigma(j)|, a, b)\) for
some \(g\), the proposition follows.
\leavevmode\unskip\penalty 9999 \hbox{}\nobreak\hfill\quad\hbox{\ensuremath{\square}}

\subsubsection*{Proof of Theorem \ref{thm:cxy}:}

The first equation follows directly from Observation 1. We have that
\(C_{a,b}(1)\) is invertible if \(a^2 \neq b^2\). From this and that
\((A \otimes B) = (A^{-1} \otimes B^{-1})\) we complete the proof.
\leavevmode\unskip\penalty 9999 \hbox{}\nobreak\hfill\quad\hbox{\ensuremath{\square}}

\subsubsection*{Proof of Lemma \ref{lem:tr}:}

From Section \ref{sec:est} we have that \begin{align*}
\operatorname{cov}(\hat{\pi}(m)) &= m^{-1} \left(C^{-1} \text{diag}(C\pi)
{C^{-1}}^T - \pi \pi^T\right). 
\end{align*} By properties of the trace of matrix products and symmetry
of \(C^{-1}\), \begin{align*}
&\operatorname{Tr}\left(
m^{-1} \left(C^{-1} \text{diag}(C\pi)
{C^{-1}}^T - \pi \pi^T\right)
\right) \\
&=
m^{-1}\left(
\operatorname{Tr}\left(C^{-1} \text{diag}(C\pi) {C^{-1}}^T\right) - \operatorname{Tr}(\pi \pi^T)
\right) \\
&= 
m^{-1}\left(
\operatorname{Tr}({C^{-1}}{C^{-1}} \text{diag}(C \pi)) - s
\right)
\end{align*} From \((A \otimes B)(C \otimes D) = (AC) \otimes (BD)\) it
follows that \(C_{a}(n) C_{a}(n) = C_{a^2 + (1-a)^2}(n)\). From this and
Theorem \ref{thm:cxy} and Corollary \ref{corr:cxy} we get that the entry
\((C^{-1}C^{-1})_{0,0} = f(n, a)\) where \[
f(n, a) = \left(\frac{a^2+(1-a)^2}{(2a-1)^2}\right)^n.
\] Furthermore, from Proposition \ref{prop:props} the diagonal entries
of \(C^{-1}C^{-1}\) are all \(f(n,a)\). Combining this, that
\(\operatorname{Tr}(AB) = \sum_{i,j}(A \odot B^T)_{i,j}\), and
\(\sum_x C_x \pi = 1\), \begin{align*}
\operatorname{Tr}(C^{-1}C^{-1} \text{diag}(C \pi)) &= 
\sum_x f(n,a) C_x \pi 
= f(n,a) \sum_x C_x \pi 
= f(n,a),
\end{align*} and consequently,
\(\operatorname{Tr}(\operatorname{cov}(\hat{\pi}_x(m))) = m^{-1} \left(f(n,a) - s\right)\).
\leavevmode\unskip\penalty 9999 \hbox{}\nobreak\hfill\quad\hbox{\ensuremath{\square}}

\subsubsection*{Proof of Theorem \ref{thm:l2formula}:}

We have that \begin{align*}
\operatorname{Tr}(\operatorname{cov}(\hat{\pi}^*)) &= m^{-1}\operatorname{Tr}(\text{diag}(\pi) - \pi\pi^T)
= m^{-1}(1 - s).
\end{align*} From Lemma \ref{lem:tr} and Proposition \ref{prop:l2} we
get that \(L= f_L(s) = \frac{c - s}{(1 - s)}\) for
\(c = \left(\frac{a^2+(1-a)^2}{(2a-1)^2}\right)^n.\) From
\(0 \leq p_x \leq 1\) and \(\sum_x p_x = 1\), \(s\) has a minimum when
\(p_x = 1/2^n\) for all \(x\), and maximum when \(p_x = 1\) for a fixed
\(x\), and \(p_y = 0\) for \(y \neq x\). These values are then
\(\frac{2^n}{(2^n)^2} = 2^{-n}\) and \(1\), respectively. The \(m\)'th
derivative of \(f_L(s) = \frac{c-s}{1-s}\) wrt. \(0 \leq s < 1\) is
\(f_L^{(m)}(s) = \frac{m!}{(1-s)^m} \left(f_L(s) - 1 \right)\). The loss
\(f_L\) therefore achieves its minimum at \(f_L(2^{-n})\).
\leavevmode\unskip\penalty 9999 \hbox{}\nobreak\hfill\quad\hbox{\ensuremath{\square}}

\subsubsection*{Proof of Proposition \ref{prop:lambda} (sketch):}

The \(m\)'th derivative of \(f_L(s) = \frac{c-s}{1-s}\) wrt.
\(0 \leq s < 1\) is
\(f_L^{(m)}(s) = \frac{m!}{(1-s)^m} \left(f_L(s) - 1 \right)\). Since
\(c \geq 1\), \(f^{(m)}_L \geq 0\) for all \(m > 0\). In particular, we
have that \(f_L\) is convex, as is \(f_L^{(m)}\) for all \(m\). Using
the expectation for a first order Taylor approximation for convex
\(f_L\) we have that for random variable \(S\) \[
f_L(\operatorname{E}(S)) \leq \operatorname{E}(f_L(S)) \leq 
f_L(\operatorname{E}(S)) + \frac{\lambda}{2} \operatorname{Var}(S)  \tag*{(*)}
\] where
\(\lambda = \max_{x \in \ensuremath{\mathcal{I}}} f^{(2)}_L(x) \geq 0\)
for suitable interval \(\ensuremath{\mathcal{I}}\). Dividing \((*)\) by
\(f_L(\operatorname{E}(S)) = L(n)\), we get \[
1 \leq \frac{\operatorname{E}(f_L(S))}{f_L(\operatorname{E}(S))} \leq 1 + \delta,
\] where \[
\delta = \frac{\lambda \operatorname{Var}(S)}{2 f_L(\operatorname{E}(S))}.
\] Let \(S = \pi^T\pi\). Recalling that \(c = c(a)^n\) and expanding
both numerator and denominator at \(n=3\) (where the minimum occurs
since \(f_L\) is increasing and \(\operatorname{Var}(S)\) and
\(\operatorname{E}(S)\) are both decreasing in \(n\)), we see that
\(\delta(n) \in O(2^{-3n})\). Applying Chebyshev's inequality, we have
that
\(P(S \geq \operatorname{E}(S) + 10 \operatorname{Var}(S)^{\frac{1}{2}}) \leq 0.01\).
Evaluating \(\delta\) at
\(\operatorname{E}(S) + 10 \operatorname{Var}(S)^{\frac{1}{2}}\) and
\(n=3\), we arrive at the numerical
bound.\leavevmode\unskip\penalty 9999 \hbox{}\nobreak\hfill\quad\hbox{\ensuremath{\square}}

\subsubsection*{Proof of Proposition \ref{prop:time}:}

Let the computation of \(Z \otimes R\) require \(t_f(n^2)\) time for
\(2\times 2\) matrix \(Z\) and \(R\) of size \(n \times n\). Then we can
compute \(R_{a,b}(n)\) at a time cost of
\(t(n) = t_f(2^{2(n-1)}) + t(n-1) = \sum_{i = 0}^n t_f(2^{2i}) = \sum_{i = 0}^n t_f(4^i).\)
Letting \(t_f(n) = k 4 n\) for some \(k\), then
\(t(n) = 4k\sum_{i=0}^n 2^{2i} = 4k \sum_{i=0}^n 4^{i} = 4k (1 + \frac{1-4^n}{1-4}) = 4k (1 + \frac{4^n-1}{3}).\)
Now we have that \(t(n) = O(4^n) = O({2^n}^2) = O(|R_{a,b}(n)|)\). In
other words, the singly recursive algorithm \(R_{a,b}(n)\) is linear in
the time in the number of elements of the output matrix as we can
perform \(t_f\) in linear time in the size of input \(R\), in fact we
can expect that the Kronecker product can be implemented with
\(k \leq 3\), due to reading, multiplication, and writing.
\leavevmode\unskip\penalty 9999 \hbox{}\nobreak\hfill\quad\hbox{\ensuremath{\square}}

\end{document}